# Enhancing Web Application Firewalls with BERT-GNN for SQL Injection Detection

**Lilliane Linnet Musoke**
Department of Computer Science
University of Reading, UK
lylliamusoke@gmail.com

**Prof. Atta Badii**
Department of Computer Science
University of Reading, UK
atta.badii@reading.ac.uk

**Ahmed Ashlam**
Department of Computer Science
University of Reading, UK
a.ashlam@pgr.reading.ac.uk

**Abstract**

Detecting sophisticated SQL Injection (SQLi) attacks remains among the most critical challenges in web applications security. This research study has resulted in an optimised hybrid BERT-GNN pipeline with improved detection accuracy and robustness while reducing false-positive and false-negative rates. SQL queries are tokenised and encoded into contextual BERT embeddings, which then initialise the node features of a Graph Neural Network (GNN) trained to classify each query, with the architecture tuned by Optuna over accuracy, precision, recall, and F1-score. The proposed model achieved 99.67% accuracy, with 99.71% precision, 99.39% recall, and 99.55% F1-score on the attack class. A sensitivity analysis, performed by perturbing graph inputs, further assessed the model robustness and yielded a low mean sensitivity score of 0.0037, indicating stable predictions under such perturbations. The results have demonstrated the potential of a novel hybrid model that couples BERT contextual understanding with the GNN structural modelling to detect sophisticated SQLi attack vectors. For open validation, the dataset, test sets and models are made available at https://github.com/mlily2024/Final-project-SQL-injection-pipeline.

## 1. Introduction and Motivation

Web applications remain the most targeted software systems, with SQL Injection (SQLi) consistently ranked among the OWASP Top 10 threats (OWASP, 2021). In an SQLi attack, an adversary injects malicious SQL code into a web application database query, which can lead to unauthorised access, data corruption, malware distribution, business disruption, intellectual property and identity theft, economic loss, cyber espionage, and full system compromise. The primary cause of SQLi, which enables adversaries to inject malicious code into user-input queries, is inadequate validation of user input (Halfond & Orso, 2005).

There is a crucial need for reliable systems to detect and prevent SQLi attacks (Trinh et al., 2023). Deployed as the first line of defence, conventional Web Application Firewalls (WAFs) rely on rule-based and signature-based approaches. While effective against known threats, they require constant updates and maintenance, which is costly and time-consuming. According to Román-Gallego et al. (2023), they also suffer high false-positive and false-negative rates and struggle to identify complex attacks. Because they depend on network administrators continually updating signatures to counter new threats, WAFs remain vulnerable to zero-day attacks; their performance relies heavily on regular maintenance and expert configuration, leaving them exposed to human error. ***Figure 1*** illustrates the SQLi attack process and its associated security risks.

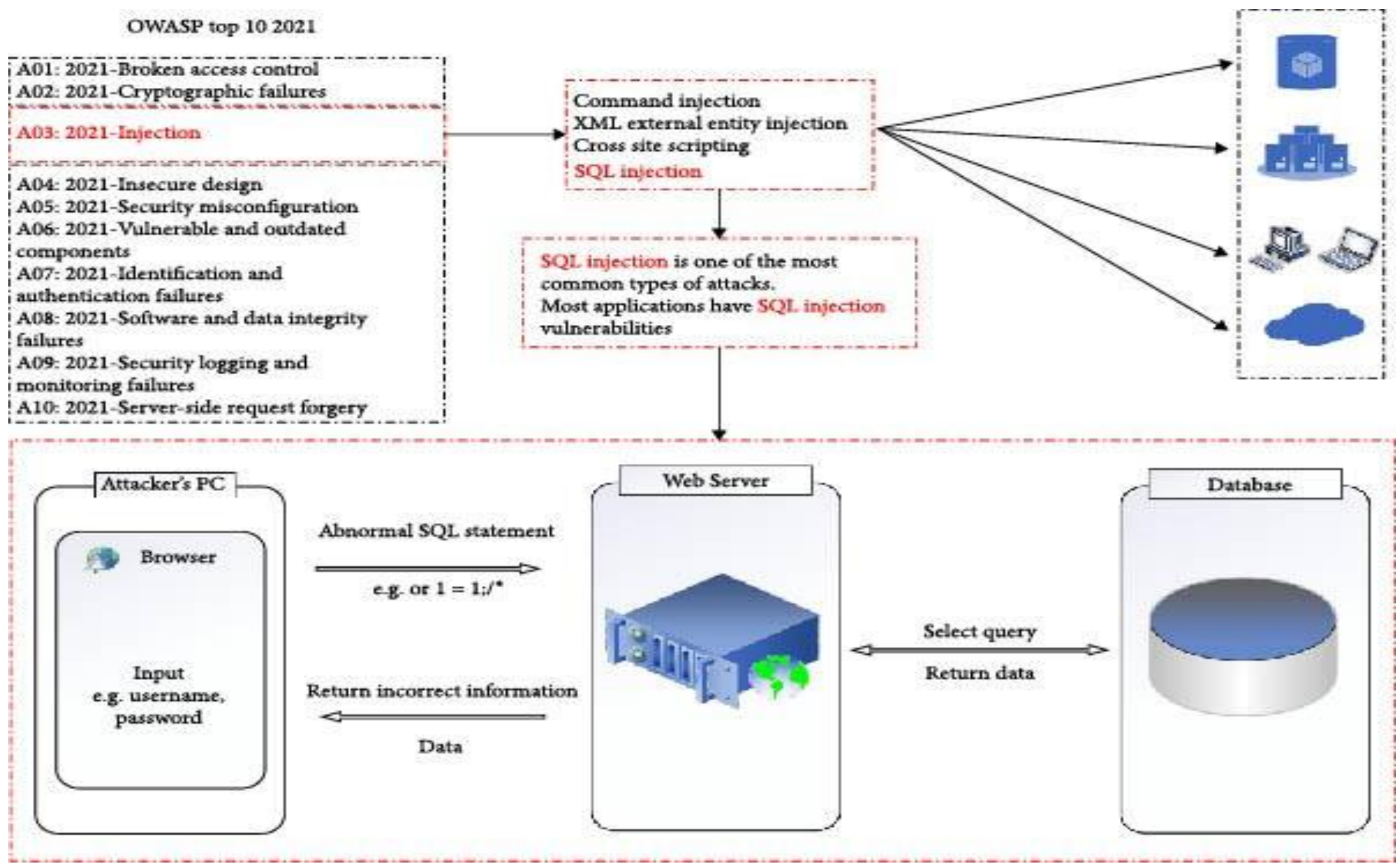

*Figure 1: Process and risks of SQLi attacks (Liu & Dai, 2024)*

**Novelty and contribution.** This paper proposes a novel hybrid model that combines Bidirectional Encoder Representations from Transformers (BERT) with Graph Neural Networks (GNN) to strengthen the security of modern web applications against evolving SQLi attacks. The hybrid pairs the deep contextual embeddings of BERT with the ability of the GNN to model graph-structured relationships between query tokens, addressing the limitations of traditional WAFs with a dynamic, adaptive, and highly accurate detection method. Crucially, and in contrast to much prior work, the model is evaluated not on accuracy alone but also on latency and on a sensitivity analysis of its robustness to perturbed inputs, the practical metrics that determine real-world deployability.

## 2. Related Work

Roy et al. (2022) used the Kaggle SQL Injection Dataset to evaluate the efficiency of Machine Learning (ML) algorithms in detecting SQL Injection (SQLi) attacks. The study executed Logistic Regression, AdaBoost, Random Forest, XGBoost, and Naive Bayes, with feature extraction based on tokenisation and payload analysis. Naive Bayes outperformed the other models, achieving the highest accuracy of 98.33%, an F1-score of 97.00%, sensitivity of 100%, and specificity of 97.71%.

Ashlam et al. (2022) developed a novel approach exploiting ML methods to detect SQLi attacks. The authors used a dataset of 1,950 malicious and 2,000 benign queries compiled from the National Vulnerability Database using the LibInjection tool. To balance the dataset and optimise performance, normal SQL queries were added, and CountVectorizer was employed for feature extraction. Support Vector Machines (SVMs), Decision Trees, K-Nearest Neighbours, AdaBoost, and Random Forests were assessed, and SVMs emerged as the best performer. Initial SVM accuracy reached 94% and, within the PALOSDM framework, improved to 99%. Iterative optimisation across accuracy, precision, recall, and F1-score demonstrated the efficiency of the approach in detecting SQLi.

Badri and Alouneh (2023) applied ML to detect malicious SQLi requests aimed at protecting cloud applications and DNS servers. The dataset used in this research was compiled from several sources, including the SQL Injection Dataset. Various preprocessing techniques cleaned and transformed the text data, with Term Frequency–Inverse Document Frequency (TF-IDF) used for feature extraction. The work assessed SVM, Gaussian Naïve Bayes, Logistic Regression, and AdaBoost. AdaBoost was the most effective, achieving 99.4% accuracy; the authors concluded that its tendency to combine weak learners and its robustness to noise provided strong protection for DNS servers and cloud applications.

Ahmed and Uddin (2020) analysed methods for detecting cyber-attacks through Natural Language Processing (NLP) and ensemble learning. The researchers collected data using open-source tools such as LibInjection and Sqlmap, producing a dataset of 36,569 self-generated SQLi payloads. Using a Bag-of-Words representation, they classified queries with Random Forest, Decision Trees, Naïve Bayes, SVM, and K-Nearest Neighbours. The Random Forest model had the highest accuracy of 98.15%, a TPR of 96.13%, an FPR of 0.02%, and an AUC of 99%.

Lu et al. (2023) proposed a semantic learning-based SQLi detection technique using datasets collected from various sources, including CVE, CNVD, the Exploit Database, and HTTP traffic samples. The authors employed synBERT to convert SQL queries into vector representations encompassing the syntactic and semantic components of SQL statements. synBERT achieved over 90% accuracy in SQLi classification, showing satisfactory performance even on unseen data.

Lakhani et al. (2022) explored the application of NLP to detecting SQLi attacks, using a dataset of 11,332 malicious and 19,588 legitimate inputs. They used NLP with BERT for feature extraction and evaluated several ML models (KNN, SVM, Gaussian Naïve Bayes, Decision Tree, and Logistic Regression). The BERT-based model achieved the highest accuracy at 97%, with a 0.8% FPR and 5.8% FNR, outperforming the traditional classifiers in the evaluation.

Liu and Dai (2024) used a BERT–LSTM hybrid for SQLi detection, employing the HTTPParams dataset of 10,852 SQL attack vectors and 19,304 normal queries. BERT provided contextual embeddings and LSTM performed classification, with feature extraction built on BERT word- and sentence-embedding generation. The hybrid recorded an accuracy of 97.3%, a precision of 96.3%, a recall of 96.2%, and an F1-score of 95.8%.

ALAzzawi (2023) studied SQLi detection using an RNN-LSTM hybrid on the Kaggle SQL Injection Dataset, which contains 30,919 instances and two features. The data was pre-processed through tokenisation and feature-extraction techniques such as CountVectorisation and TF-IDF. The research applied hybrid Recurrent Neural Network models (LSTM and GRU): the RNN-LSTM model achieved the best accuracy of 90.12%, while the RNN-GRU model achieved 84.89%. The author concluded that the hybrid models detected SQLi attacks more effectively than rule-based methods.

Gandhi et al. (2021) introduced a CNN-BiLSTM hybrid to detect SQLi attacks, using a dataset of 4,200 queries (3,072 malicious and 1,128 normal). After tokenisation, HTTP requests were denoised and decoded to obtain word embeddings via Word2Vec, and a CNN-MLP hybrid distinguished malicious queries. In a comparative study, the CNN-BiLSTM hybrid achieved 98% accuracy, surpassing the standalone ML algorithms evaluated.

Collectively, these studies demonstrate tremendous progress in SQLi detection but also share several limitations. Many rely on limited or self-generated datasets, potentially restricting the generalisability of their findings. While reported accuracy rates are high, insufficient attention is often paid to other crucial metrics such as false-positive rates, adversarial accuracy, latency, and sensitivity analysis. A further common limitation is a lack of adaptability to evolving attacks: most models presented are static, making them vulnerable. These gaps underscore the need for research into advanced approaches that combat obfuscated, zero-day attacks while enhancing detection accuracy, robustness, and generalisability in the real world, the motivation for the BERT-GNN model proposed here.

## 3. Methodology

The architecture leverages BERT for natural language understanding and a GNN to capture the structural relationships between tokens in a query. BERT, a transformer-based model with deep learning and NLP capabilities, processes an entire sentence in both directions to capture contextual relationships between words (Devlin et al., 2018). The BERT tokeniser divides each SQL query into tokens, producing token IDs and contextual embeddings that represent the semantic meaning of the tokens in relation to the query. These 768-dimensional embeddings serve as the node features for the GNN.

### 3.1 Graph construction

Each SQL query is represented as a graph $G = (V, E)$, where the nodes $V$ are the query tokens and the edges $E$ encode the relationships between them. The graph is built as a **sequential chain (a path graph)**: each token node is connected to the immediately following token, with edges undirected (added in both directions, consistent with the symmetric normalisation of Equation 6), preserving the left-to-right order of the query. It is not a fully connected graph and does not use similarity-thresholded edges; token i is linked only to token i+1. Each node is initialised with the 768-dimensional BERT embedding of its token. The graph is constructed with NetworkX and converted to a PyTorch Geometric (PyG) data object for training.

### 3.2 GNN architecture

The GNN comprises **two graph convolutional layers (GCNConv)** (Kipf & Welling, 2017), each followed by a ReLU activation, which aggregate information from neighbouring nodes to produce representations that combine structural and token-level information. A **global mean pooling** operation then aggregates the node features into a single feature vector representing the entire query. This vector passes through a **dropout layer** and **two fully connected layers: the first** (hidden→hidden) with a ReLU activation, the second (hidden→2) producing the class logits. This design leverages both BERT language representation strengths and the GNN structural relationship-capturing strengths, making SQLi detection more robust and comprehensive by capturing contextual and relational information in each query.

The mathematical formulation of the pipeline is as follows.

**Token Embedding:** convert each token in the SQL query into a fixed-size vector representation.

$$Token\ Embedding: E = [e_1, e_2, \dots\dots, e_n] \tag{1}$$

$$where\ e_i\ is\ the\ embedding\ for\ the\ i-th\ token\ and\ n\ is\ the\ number\ of\ tokens.$$

**Position Embedding:** Add positional information to each token embedding

$$E_{pos} = E + P \tag{2}$$

Where P is the positional embedding matrix.

**Attention Mechanism:** Calculate attention scores to capture the contextual relationships between tokens.

$$Attention: Attention(Q, K, V) = softmax\left(\frac{QK^T}{\sqrt{d_k}}\right)V \tag{3}$$

Where Q(queries), K(keys), and V(values) are linear transformations of the input embeddings.

**Self-attention (internal to BERT): BERT applies multiple self-attention layers internally to refine token representations; Equations 3 and 4 describe this operation of the pretrained encoder, not layers implemented in this pipeline.**

$$H^{(l+1} = layernorm\left(H^l + Attention\left(H^{(l)}\right)\right) \quad (4)$$

$$where\ H^{(l)}\ is\ the\ hidden\ state\ at\ layer\ l$$

**GNN: Structural Analysis**

**Graph Construction**: Represent the SQL query as a graph G = (V, E), Where V are the nodes (tokens), and E are the edges (relationships between tokens).

**Node Embedding Initialisation**: Initialise node features using BERT embeddings.

$$X = [x_1, x_2, \ldots., x_n] \quad (5)$$

$$where\ x_i\ is\ the\ BERT\ embedding\ for\ node\ i$$

**Graph convolution**: Aggregate information from neighbouring nodes.

$$h_i^{l+1} = \sigma\left(\sum_{j \epsilon N(i)} \frac{1}{c_{ij}} W^{(l)} h_j^{(l)}\right) \quad (6)$$

$$where\ h_i^{(l)}\ is\ the\ hidden\ state\ of\ node\ i\ at\ layer\ l, N(i) is\ the\ neighborhood\ of\ node\ i, c_{ij}\ is\ a$$

$$normalisation$$

$$constant, W^{(l)}\ is\ the\ learnable\ weight\ matrix, and\ \sigma\ is\ an\ activation\ function.$$

**Graph Pooling**: Aggregate node representations to form a graph-level representation.

$$h_G = pool\left(\left\{h_i^{(L)} \mid i \in V\right\}\right) \quad (7)$$

Where L is the number of layers and pool is a pooling function (e.g., global mean pooling).

**Final Classification**

**Graph-level representation for classification: the GNN node features, initialised from the BERT contextual embeddings, are mean-pooled into a single graph vector that is passed to the fully connected classifier.**

$$z = h_G \quad (8)$$

$$where\ hG\ is\ the\ pooled\ GNN\ representation\ passed\ to\ the\ classifier.$$

**Classification Layer**: Use a fully connected layer to predict the class label.

$$\hat{y}_i \hat{y} = softmax(W_c z + b_c) \quad (9)$$

$$where\ W_c\ and\ b_c\ are\ learnable\ parameters\ of\ the\ classification\ layer.$$

**Loss Function**

**Cross-Entropy Loss**: Compute loss between predicted labels and true labels.

$$L = -\sum_i w_i y_i \log(\hat{y}_i) \quad (10)$$

$$where\ yi\ is\ the\ true\ label, \hat{y}i\ is\ the\ predicted\ probability\ for\ class\ i, and\ wi\ is\ the\ balanced\ class\ weight\ for\ class\ i \hat{y}_i$$

This dual methodology makes the model resilient to diverse SQLi attacks by efficiently representing both the structure and the content of SQL queries.

## 4. Implementation

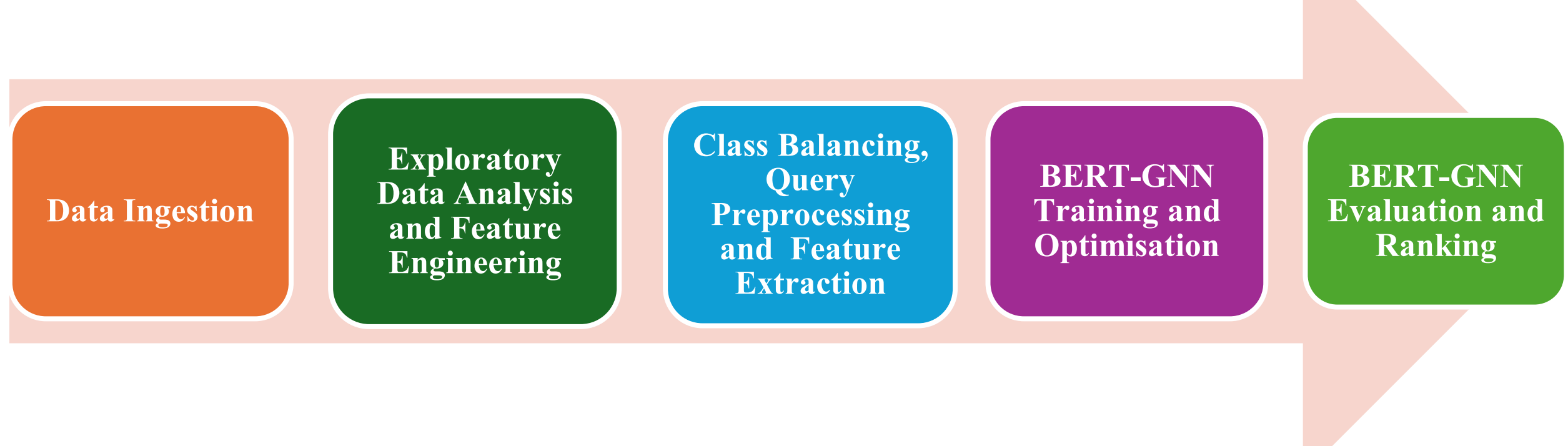

***Figure 2:*** *BERT-GNN pipeline implementation workflow, from data ingestion to model evaluation and ranking.*

### 4.1 Data ingestion

The first step was to import the dataset; the study used the SQL Injection Dataset from Kaggle, which contains 30,919 instances classified as malicious or benign, with two features (Query and Label). The dataset was loaded into a Pandas DataFrame to facilitate manipulation and analysis. Its dimensions, shape, descriptive statistics, and data types were examined to understand the data format and properties, informing the preprocessing steps.

### 4.2 Exploratory data analysis and feature engineering

Several cleaning steps prepared the data for analysis. Queries were converted to lowercase and stripped of leading and trailing whitespace to remove inconsistencies from case sensitivity and unnecessary spacing. The dataset was checked for missing values and found to have none. A new feature, Query_Length, was created to represent the number of words in each query, aiding the understanding of query distribution and characteristics. The 99th percentile for query length was 79 words; no data points were removed to retain the dataset completeness and include all query variations, including outliers. Target labels were encoded with scikit-learn LabelEncoder to convert categorical labels into the numerical format required by the models. Finally, the dataset was split into training and test sets using a 70%–30% split, so the model was trained on most of the data while retaining a portion for a reliable measure of its generalisability to unseen data.

### 4.3 Class balancing, query preprocessing, and feature extraction

**Class balancing.** The dataset contained a significant class imbalance (***Figure 3***): 63.2% normal web queries and only 36.8% SQLi attacks. Balanced class weights were computed for the BERT-GNN model, assigning greater weight to the minority class during training so that both classes contributed comparably to the loss, improving the ability of the model to identify minority-class instances.

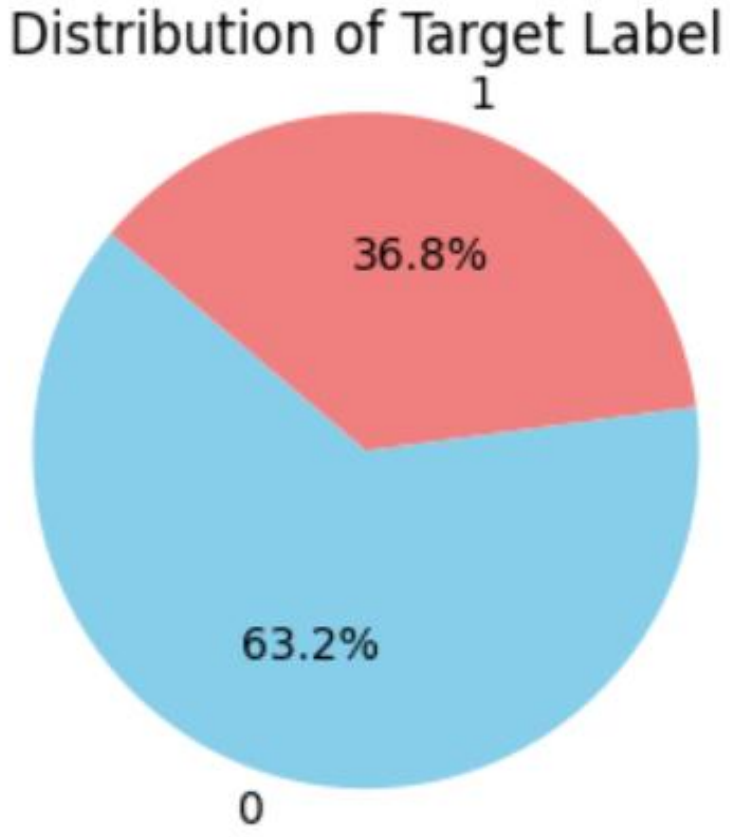

***Figure 3:*** *Distribution of the target variable.*

**Data preprocessing.** Tokenisation was performed by passing the SQL queries through the BERT tokeniser, which produced token IDs. Each query was split into tokens, [CLS] and [SEP] markers were added, and each token was assigned an integer ID. Sequences were padded (shorter queries) or truncated (longer queries) to a standard length of 128 tokens, so all tokenised sequences had equal length. For each query, a sequential-chain graph was constructed with tokens as nodes and adjacent-token connections as edges, built with NetworkX and converted to a PyG data object. These procedures put the data in a suitable format for training the hybrid BERT-GNN model.

### 4.4 Training procedure and hyperparameter optimisation

The BERT-GNN model was tuned using the hyperparameter optimisation library Optuna, which searched over batch size, learning rate, dropout rate, and hidden layer size across 40 trials. Early stopping (patience of 5 epochs, maximum 50 epochs) prevented overfitting by terminating training when the validation loss did not improve; the run stopped after twelve epochs (Figure 5). The model used the Adam optimiser to update parameters from the computed gradients and a class-weighted Cross-Entropy loss for classification. Because the dataset was large, training was performed using mini-batches, which reduced memory pressure and helped stabilise training. The best hyperparameters found by Optuna are given in ***Table 1***.

**Table 1: Best BERT-GNN hyperparameters (Optuna, 40 trials)**

| Hyperparameter | Search range | Best value |
|---|---|---|
| Hidden dimension | 32–256 | 100 |
| Dropout rate | 0.1–0.5 | 0.139 |
| Learning rate | 1e-5–1e-2 | $1.13 \times 10^{-3}$ |
| Batch size | 32–128 | 74 |
| Optimiser | N/A | Adam |
| Loss | N/A | Class-weighted Cross-Entropy |

### 4.5 Model evaluation and ranking

The model was evaluated on accuracy, precision, recall, and F1-score. Accuracy is the ratio of correctly predicted instances to all instances. Precision, recall, and F1-score assessed the model effectiveness at identifying positive (malicious) and negative (benign) instances: precision measured the accuracy of the positive predictions, recall evaluated the model capacity to identify all positive instances, and the F1-score combined the two into a balanced measure. The confusion matrix provided a breakdown of true positives, true negatives, false positives, and false negatives.

Resource utilisation was also considered, focusing on the model computational efficiency in terms of latency, to assess the practical viability of real-world deployment. Finally, to evaluate robustness, a sensitivity analysis was performed on the BERT-GNN model by perturbing nodes in the input graphs and observing the change in predictions; the average change was used to compute a sensitivity score, indicating the degree to which the model is susceptible to changes in its input structure.

### 4.6 Experimental platform

All experiments were carried out in Google Colab Pro on an NVIDIA Tesla T4 GPU runtime. The pipeline was implemented in Python 3 with PyTorch and PyTorch Geometric for the graph model, the Hugging Face Transformers library for the pretrained BERT-base-uncased encoder, NetworkX for graph construction, Optuna for hyperparameter optimisation, and scikit-learn for evaluation. All execution times reported in this paper were measured in this environment. Execution time is hardware-dependent and is not directly comparable across platforms, whereas accuracy is platform-independent.

## 5. Results

The BERT-GNN results, summarised in **Table 2,** indicate robust performance in detecting SQLi attacks. The model achieved an accuracy of 99.67%, with attack-class precision, recall, and F1-score of 99.71%, 99.39%, and 99.55% respectively, a total execution time of 23.13 seconds (as run on the computing platform in Section 4.6), and a mean sensitivity score of 0.0037.

**Table 2: Performance metrics - BERT-GNN (accuracy over both classes; precision, recall and F1-score for the malicious/attack class)**

| Model | Accuracy | Precision | Recall | F1-score | Execution time (s) | Sensitivity |
|---|---|---|---|---|---|---|
| BERT-GNN | 0.9967 | 0.9971 | 0.9939 | 0.9955 | 23.13 | 0.0037 |

**Figure 4** highlights the model effectiveness at minimising false positives and false negatives: it correctly classifies the great majority of benign and malicious queries, with only 10 false positives and 21 false negatives. This low misclassification rate is crucial in SQLi detection, minimising both the likelihood of missing genuine attacks (false negatives) and of flagging benign queries as malicious (false positives).

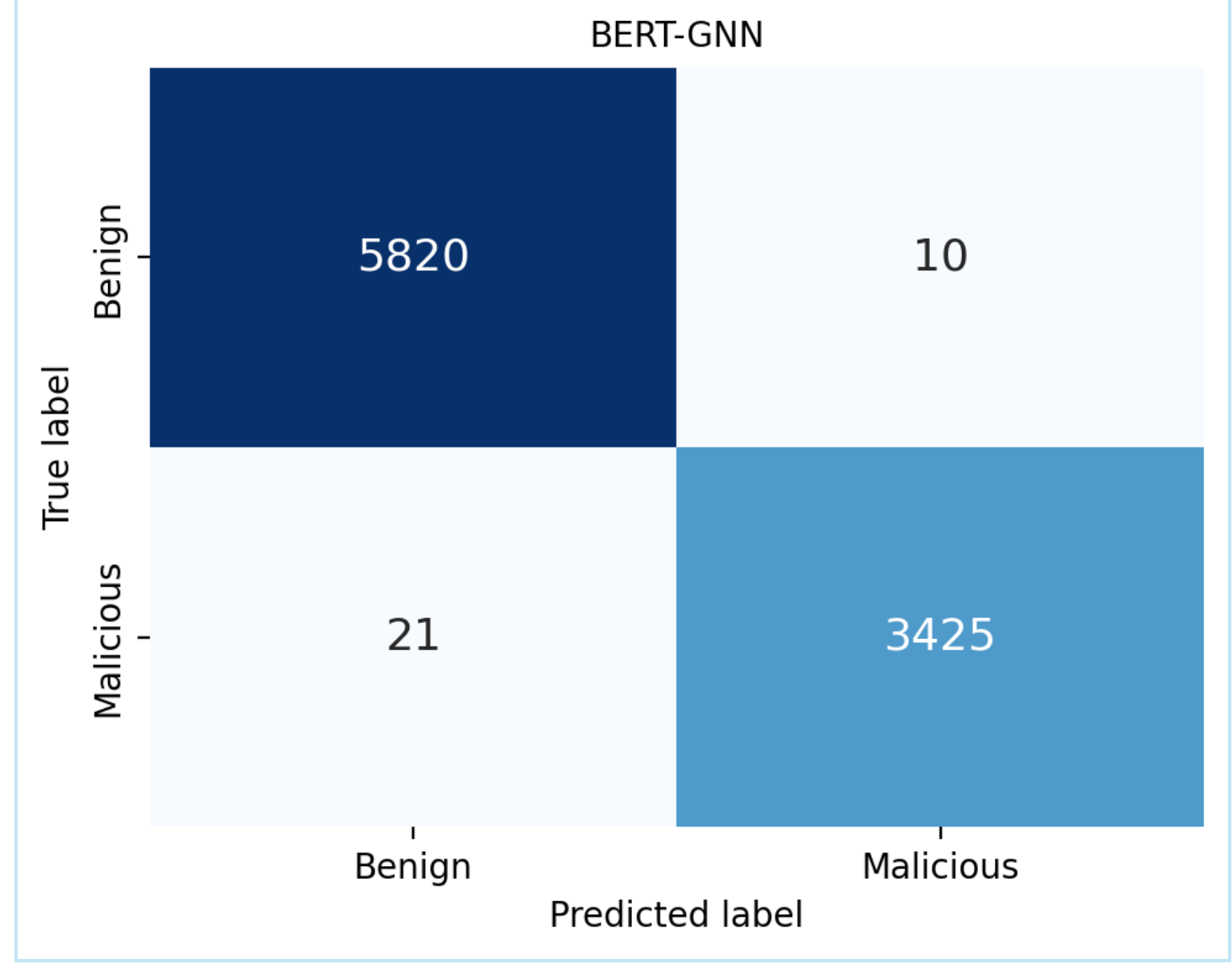

***Figure 4:*** *Confusion matrix - BERT-GNN.*

**Figure 5** shows the held-out BERT-GNN training and validation loss over the twelve epochs before early stopping. The training loss falls sharply from about 0.055 to below 0.01 within the first few epochs and continues to decline to roughly 0.001, while the validation loss settles to around 0.01 and fluctuates within a narrow band between 0.008 and 0.022. The small and stable gap between the two curves indicates that the model converges quickly and generalises well, without overfitting. Figure 6 shows a near-perfect Area Under the Curve (AUC) of 1.00, indicating a strong ability to distinguish between benign and malicious SQL queries with high sensitivity and specificity.

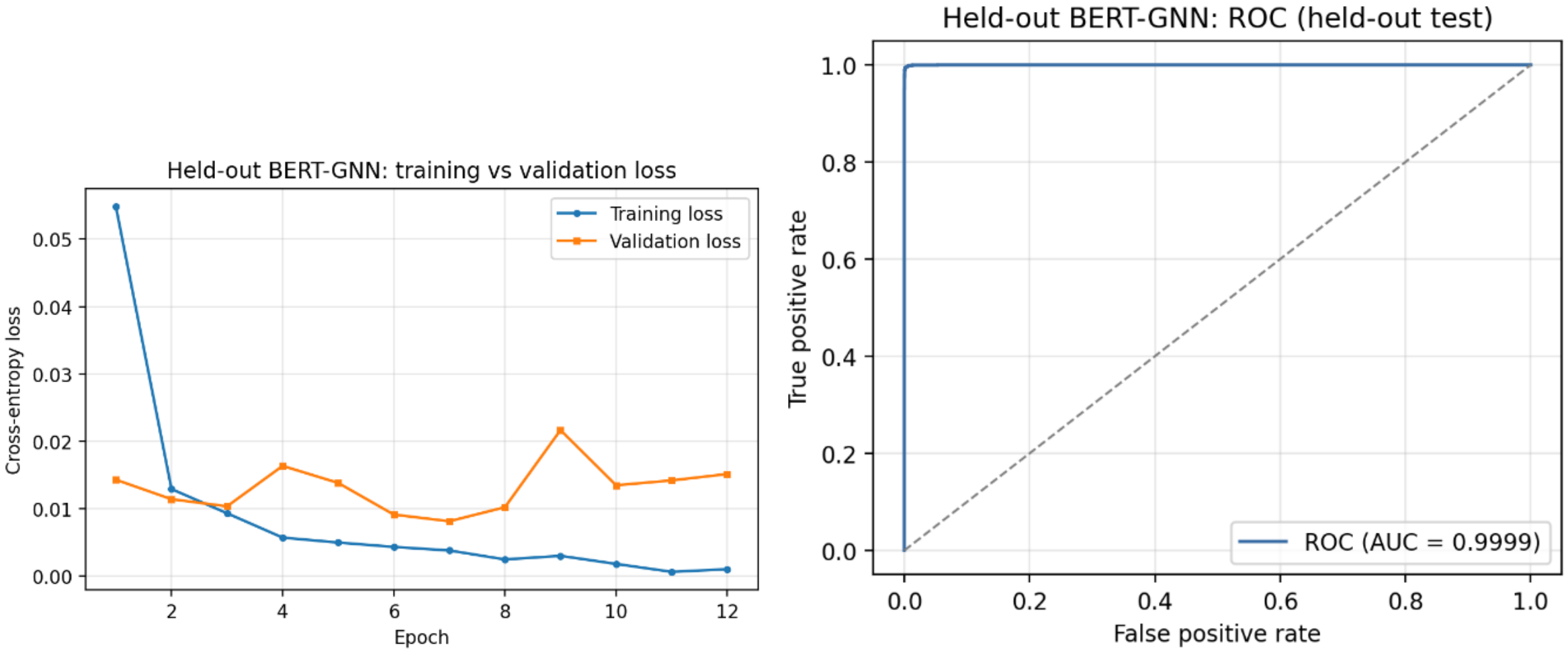

*Figure 5: Training and validation loss over epochs.*

*Figure 6: Receiver Operating Characteristic (ROC) curve - BERT-GNN.*

**Figures 7** and ***8*** examine the misclassifications. A range of query lengths can cause some benign queries to be flagged (**Figure 7**), indicating the need for additional contextual understanding or feature refinement. The false negatives occur primarily at shorter query lengths (***Figure 8***), implying that certain concise attack vectors can bypass detection and highlighting an area for improvement in the model sensitivity to short, malicious queries.

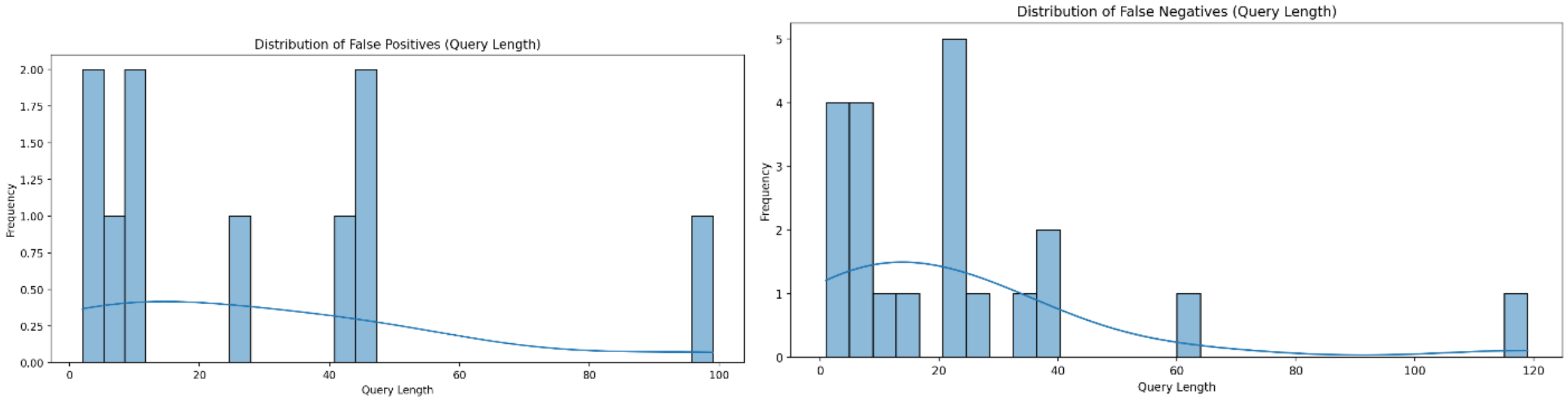

*Figure 7: Distribution of false positives - BERT-GNN.*

*Figure 8: Distribution of false negatives - BERT-GNN.*

## 5.1 Sensitivity analysis

The held-out BERT-GNN averaged a sensitivity score of 0.0037 across 99 malicious attack-pattern graphs, indicating a very low overall susceptibility of the model predictions to changes in graph structure. Sensitivity here is the mean absolute change in the predicted malicious-class probability when about 10% of the nodes of a graph are removed at random. The scores are strongly skewed towards zero: 98% of the graphs record a sensitivity below 0.01 and the median is effectively zero, so for almost every attack pattern the prediction is unchanged by node removal. A single pattern stands out as more fragile, graph 86 (0.307), where removing a critical token shifts the prediction, while every other graph stays below 0.03. Overall, the model is highly stable, and the one high-sensitivity case pinpoints an attack pattern that depends on an individual token.

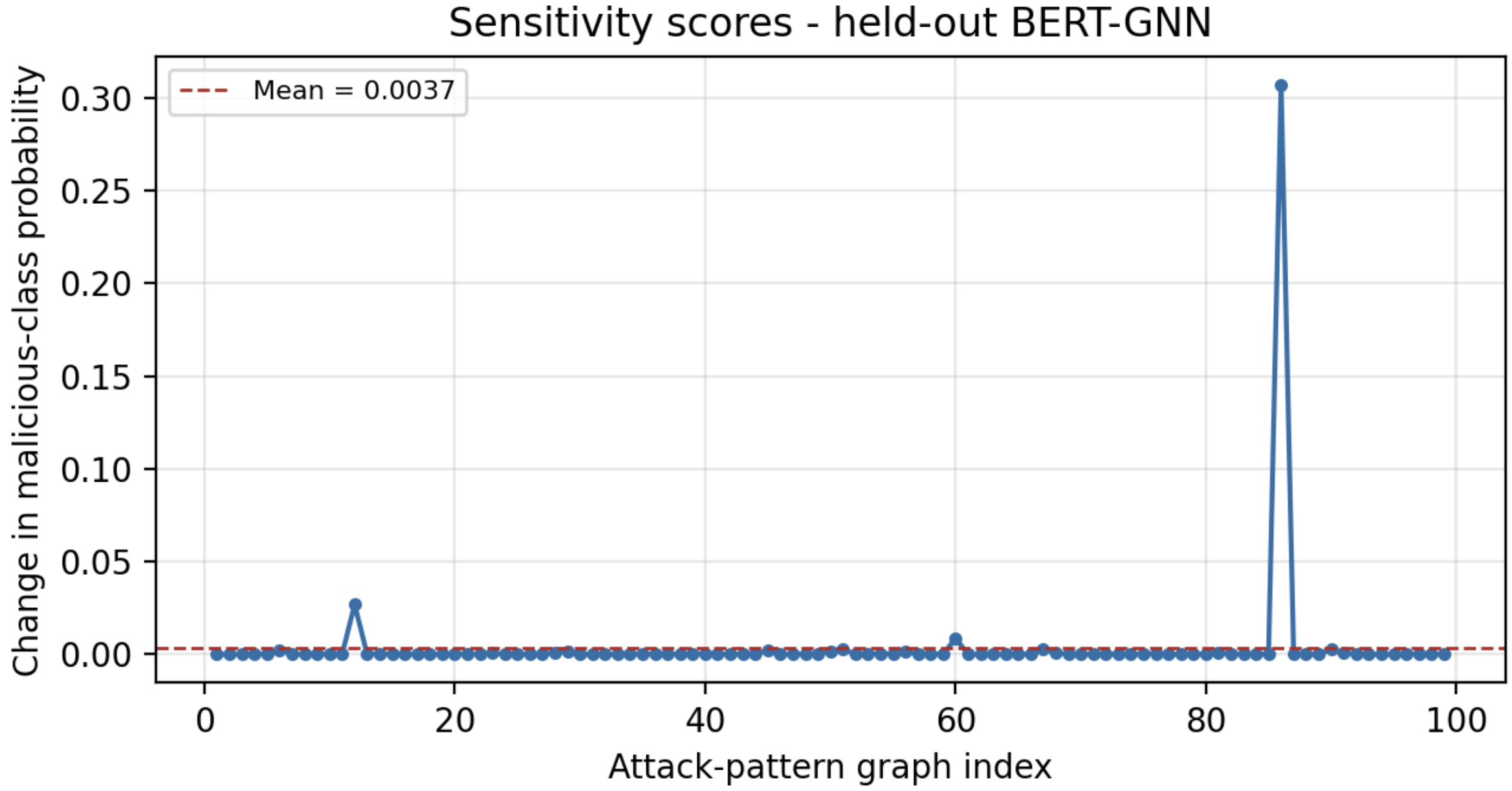

***Figure 9:*** *Sensitivity scores - BERT-GNN.*

**Figure 9** visualises the sensitivity scores across the 99 attack-pattern graphs. Almost all scores sit at or near zero, confirming that the prediction is stable across the attack patterns when about 10% of the input nodes are removed. A single isolated peak, reaching about 0.31, marks the one attack pattern where the prediction depends on a small number of critical tokens, highlighting a specific case to strengthen for improved reliability in real-world deployment.

### 5.2 Isolating the contribution of BERT and the GNN

To isolate the contribution of the GNN component, we trained a BERT-only baseline on the identical 70/30 split: the contextual [CLS] embedding of BERT fed to a simple classification head, with no graph and no GNN. On the identical 9,276-query test set the BERT-only baseline reaches 99.66% (Logistic Regression) and 99.73% (a single-hidden-layer MLP), within roughly 0.07 points of the BERT-GNN hybrid at 99.67% (Table 3), so the three are effectively level. The clean-accuracy performance is therefore attributable to the BERT encoder; the graph adds no meaningful accuracy on this near-saturated benchmark.

**Table 3. BERT-only ablation on the identical 9,276-query test set: the graph adds no clean accuracy.**

| Model | Accuracy (%) | Precision (%) | Recall (%) | F1 (%) |
|---|---|---|---|---|
| BERT-GNN hybrid | 99.67 | 99.71 | 99.39 | 99.55 |
| BERT [CLS] + Logistic Regression | 99.66 | 99.68 | 99.39 | 99.54 |
| BERT [CLS] + MLP | 99.73 | 99.80 | 99.48 | 99.64 |

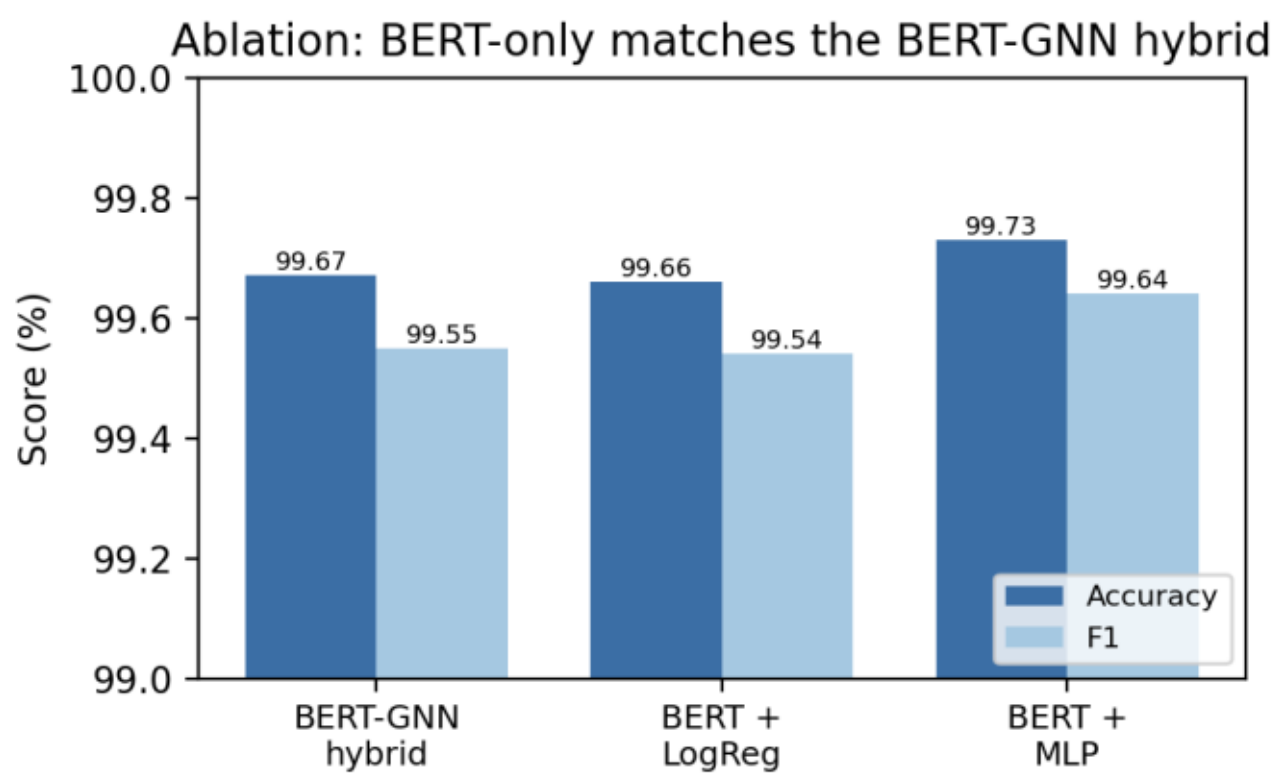

***Figure 10****. Ablation: a BERT-only baseline matches the BERT-GNN hybrid on the identical test set.*

To test whether the graph structure itself carries any signal, we retrained the model with its edges removed entirely, and again with random edges of the same count. Test accuracy is unchanged, 99.65% with no edges and 99.59% with random edges, against 99.67% with the real structural edges, all within seed variation, so the graph topology is inert on this task: the model classifies by pooling the BERT token features, a role the [CLS] summary fulfils equally well.

The graph is not, however, without value. Under Gaussian perturbation of the input features, evaluated across five random seeds, the GNN flips only 0.62% (plus/minus 0.03) of its predictions at a moderate noise level and 3.42% (plus/minus 0.30) at a higher one, against the 10.04% and 24.85% of the BERT-only baseline **(Table 4), a large and reproducible** robustness to input corruption. This follows from the model aggregating over many token representations, so that per-token noise averages out, whereas the single-vector baseline is perturbed directly. It is the one property on which the graph model reliably surpasses BERT alone.

**Table 4. Multi-seed robustness (five seeds): of the properties tested, only robustness to random feature-noise is a reproducible GNN advantage; the evasion and OOD advantages seen in single runs were seed artefacts.**

| Robustness property | Structure-GNN | BERT-only | Reproducible advantage |
|---|---|---|---|
| Feature-noise flip, sigma 0.5 (%) | 0.62 +/- 0.03 | 10.04 | Yes |
| Feature-noise flip, sigma 1.0 (%) | 3.42 +/- 0.30 | 24.85 | Yes |
| URL-encode recall (%) | 90.08 +/- 9.48 | 90.34 | No |
| Adaptive best-of-N evasion (%) | 10.15 +/- 9.26 | 8.00 | No |
| OOD false-positive rate (%) | 7.64 +/- 3.35 | 11.76 | No |

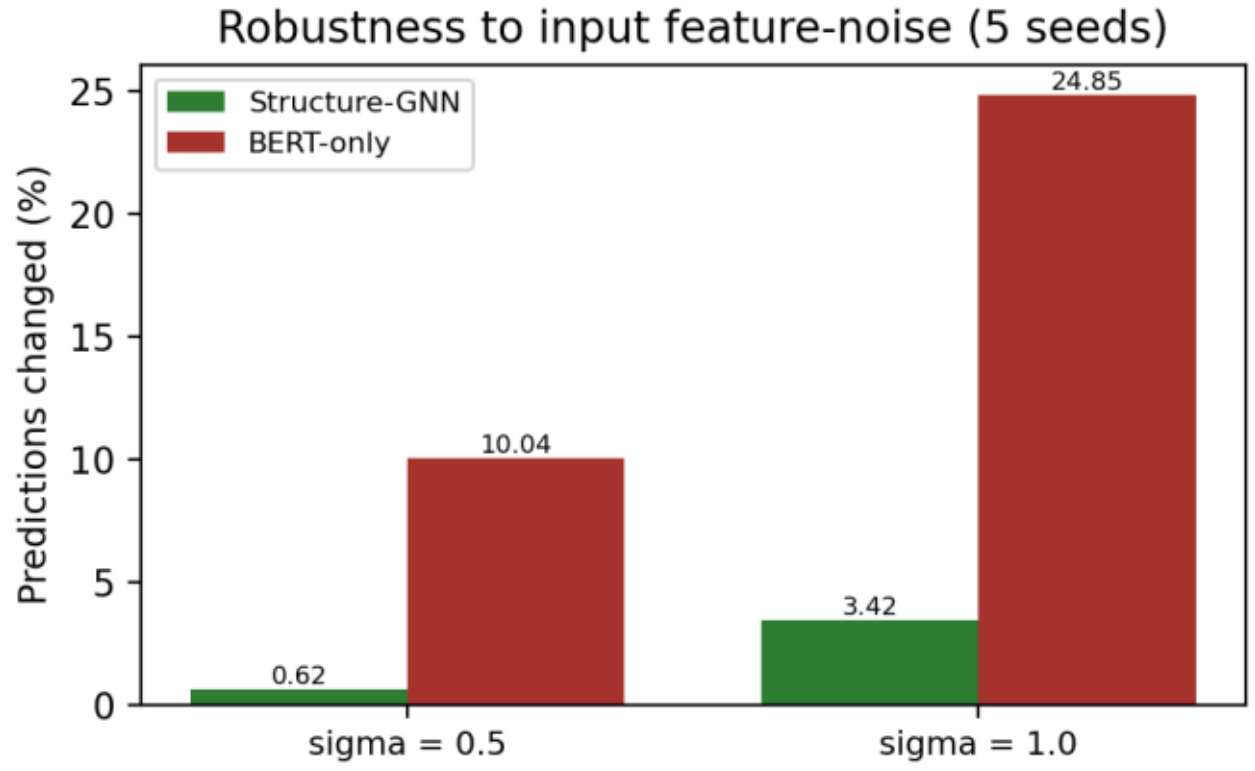

***Figure 11.*** *Robustness to input feature-noise (five seeds): the GNN changes far fewer predictions than BERT-only.*

We evaluated further robustness properties, detection under URL-encoding and adaptive evasion, and out-of-distribution (OOD) false alarms, but under the same five-seed protocol only the feature-noise robustness proved reproducible; the single-run advantages on the other metrics were artefacts of the random seed (Table 4), and under worst-case white-box gradient attacks the GNN is the more vulnerable of the two models.

In summary, on this benchmark BERT supplies the discriminative accuracy and the GNN supplies robustness to input-feature corruption, a narrow but genuine division of labour that a clean-accuracy comparison alone could not reveal.

## 6. Conclusions

This study has delivered an optimised hybrid BERT-GNN model to enhance the detection and mitigation of SQLi attacks on web applications within WAFs. The proposed approach achieved 99.67% accuracy (99.55% attack-class F1-score), and a low mean sensitivity score of 0.0037 confirmed its robustness to input perturbations. By pairing BERT contextual embeddings with the GNN structural modelling, the approach has shown improvements in accuracy, robustness, and scalability with an enhanced capability to detect complex, novel SQLi attacks. For open validation, the dataset, test sets and models are made available at https://github.com/mlily2024/Final-project-SQL-injection-pipeline.

## 7. Future Work

The experiments reported here point to several evidence-driven directions. Because the benchmark is near-saturated, with a BERT-only baseline matching the hybrid and edge ablation showing the token-chain topology to be largely inert, future work should move onto harder, non-saturated data drawn from real web-application-firewall traffic and diverse payload sources, with cross-dataset training and testing to measure genuine generalisation rather than in-distribution performance. To make the graph itself contribute, the linear chain could be replaced with semantically richer structures such as abstract syntax, parse or data-flow graphs with typed edges, alongside architectures beyond the graph convolutional network used here, including graph attention networks, GraphSAGE, graph isomorphism networks and graph transformers. Robustness should be extended from the random feature-noise the model already resists to adaptive adversaries through adversarial and certified-robustness training, while the residual false negatives at short query lengths could be tackled with hard-example mining, focal or cost-sensitive losses and short-payload augmentation. Finally, lighter encoders such as DistilBERT, ALBERT or TinyBERT with distillation and quantisation would support real-time deployment, and, given that several single-run advantages here vanished across seeds, multi-seed statistically tested evaluation should become standard practice.